# Observation of Gain Spiking and Nonlinear Beating of Optical Frequency Comb in a Microcavity

Yuanlin Zheng,[1] Tian Qin,[2] Jianfan Yang,[2] Xianfeng Chen,[1*]
Li Ge[3*] and Wenjie Wan[1,2*]

[1]The State Key Laboratory of Advanced Optical Communication Systems and Networks,
Department of Physics and Astronomy, Shanghai Jiao Tong University, Shanghai 200240, China
[2]The University of Michigan-Shanghai Jiao Tong University Joint Institute,
MOE Key Laboratory for Laser Plasmas and Collaborative Innovation Center of IFSA,
Shanghai Jiao Tong University, Shanghai 200240, China
[3]Department of Engineering Science and Physics, College of Staten Island,
the City University of New York, NY 10314, USA
*Correspondence and requests for materials should be addressed to:
Wenjie Wan (wenjie.wan@sjtu.edu.cn), Li Ge (li.ge@csi.cuny.edu) or Xianfeng Chen (xfchen@sjtu.edu.cn)

**Optical frequency combs are crucial for both fundamental science and applications demanding wide frequency spanning and ultra-precision resolutions. Recent advancements of nonlinear Kerr effect based optical frequency combs in microcavities open up new opportunities in a compact platform, however, internal cavity-enhanced nonlinearities are still unclear. Here we demonstrate transient nonlinear dynamics during optical frequency comb formation inside a Kerr microcavity. We show that gain spiking forms due to nonlinear phase modulation, causing comb lines' self-detuning nearby a cavity resonance, this introduces one key mechanism to stabilize optical frequency combs. Moreover, nonlinear beating has be observed by injecting an external probe to exam nonlinear cross-phase modulation between comb lines. Nonlinear transient dynamics here reveal the hidden nonlinear features of Kerr based optical frequency combs, leading to a new direction for ultrawide, ultrastable and ultrafast frequency comb generation in microcavities.**

Optical frequency combs (OFCs), combining thousands of ultrasharp laser lines over a wide frequency spectrum, enable ultrashort laser pulsing for important applications in optical metrology [1], precision spectroscopy [2], microwave generation [3,4], optical clock [5]. Traditional OFCs based on mode-locked femtosecond laser technology such as Ti:sapphire solid state lasers or Er/Yb doped fiber lasers, are bulky and lack of robustness. Recent advancements of OFCs based on nonlinear Kerr effect in microcavities have promising futures for compact and stable comb generations [6-9]. Unlike frequency combs in conventional optical gain media, such OFCs in microcavities rely on nonlinear Kerr effects to provide parametric gains. These nonlinear gains usually company with other nonlinear processes complicating the comb generations, e.g. Raman lasing[10,11], Brillouin scattering [12,13], modulation instability[14], soliton[15-18], which have been yet well studied. However, Kerr-related nonlinearities are as important as the original parametric gain regulating comb generations: like its counterpart in lasers, nonlinear gain saturation, the main mechanism limiting multimode lasing, can also cause laser spiking effect [19], which finally leads to the idea of mode-locked pulse laser. Here these important nonlinear processes will eventually play a major role in optical comb generation in microcavities, especially in the transient regime, where the next emerging area will explore the possibility of experimenting mode-locking/pulsing lasers in such compact forms [17,18, 20]. On the other hand, OFCs form on parametric four-wave mixing (FWM) processes, making it hard to break this spectrum symmetry around the central axis. But there also exist demanding quests to produce asymmetric OFCs, i.e. comb lines asymmetrically grow on one spectrum side than the other one [21], to directionally expand comb lines into particular spectrum region which are technically difficult to reach, e.g. mid-infrared, extremely ultraviolet [22]. We may explore additional freedom such as Kerr nonlinearity to assist this quest. Hence, it is essential to understand how these nonlinear processes behave during optical frequency comb generations in microcavities.

In this work, we experimentally observe transient nonlinear dynamics, i.e. gain spiking and nonlinear beating, during optical frequency comb formations through four-wave mixing inside a Kerr based whispering-gallery mode (WGM) microcavity. We show that gain spiking forms when nonlinear phase modulation plays a major role in self-detuning the nearby cavity resonances, effectively regulating gain and loss balance for each comb line. This self-stabilization is crucial in this cavity-enhanced gain process, especially in the transient regime. Moreover, this nonlinear phase modulation can also lead to nonlinear beating between an external probe and comb lines, further verifying its stabilization feature. Such nonlinear phase modulation may be the nonlinear root for asymmetric comb generation. We expect the current discoveries may uncover the importance of Kerr-related nonlinearities beside FWM gains during OFC generation, will also be beneficial for future applications in ultrafast optics, wide spectrum generation and ultra-precision metrology in a compact platform.

OFCs arise from Kerr-enabled FWMs inside a WGM microcavity: this cavity-enhanced parametric process simultaneously converts two incoming pump photons into one "signal" and another "idler" as shown in Fig. 1. When the intensities of the generated photons build up, they can successively cascade into next generations of photon pairs to form the "comb". In the meantime, rigid phase matching conditions must be met to fulfill the momentum conservation. However, such phase matching conditions not only contain linear factors, e.g. mode dispersion, material dispersion [23], but also nonlinear ones. The most influential one is the pump-induced self-phase modulation (SPM) to itself and the cross-phase modulation (XPM) to signals and idlers [9,24]. These nonlinear phase modulations (NPM, including SPM and XPM) effectively alter the effective path lengths photons undergo inside the cavity, changing their resonances. Consequently, this nonlinearity highly sensitive to pumps' intensity leaves the narrow cavity-enhanced gains fluctuating around the threshold level against the cavity losses, making OFCs unstable. Similar situations occur in conventional lasing processes where nonlinear gain saturation can cause the instability of laser operation, letting spiking or oscillations dominant during transient regime [19]. A proper designed frequency locking mechanism is required to ensure stable OFC generations, e.g. thermal locking[25]. Moreover, along with growing power of pumps, the succeeding generations of signals and idlers in comb line series start to trigger SPMs and XPMs to each other, limiting the continuum expansion of comb line spectrum, or initiating a new family of comb line of their own [26]. This poses the gain competitions between different comb families, much like mode competitions in multimode lasers, causing instability and making it hard to achieve self-frequency locking to stabilize OFCs [27]. Previously, these NPMs were totally under-looked, until recently, the increasing demands to push ultrashort pulsing with OFCs in compact microcavities, these NPMs can no long be omitted.

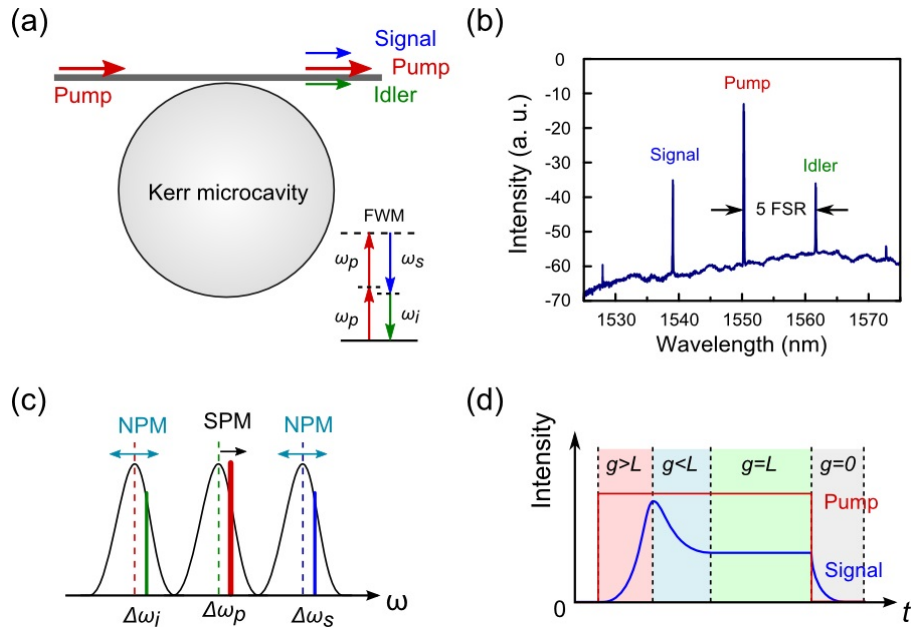

Figure 1. Illustrations of nonlinear phase modulations and gain spiking of Kerr frequency comb generation in a WGM resonator. (a) OFCs in a taper fiber coupled silica microsphere cavity. (b) The experimentally observed frequency comb spectrum. The pump wavelength is 1550.60 nm. (c) Nonlinear phase modulation's self-detuning mechanisms near cavity resonances, effectively regulating gain and loss balance for comb lines. (d) The schematic illustration of the transient dynamics of OFC generation, cavity-enhanced gain *vs.* loss.

More physics insights can be gained from the coupled rate equations (1) and (2) that govern the parametric FWM processes during OFCs in microresonators. Here the external injected strong pump $A_p$ through degenerate parametric FWMs converts energy to signal and idler beams $2\omega_p = \omega_s + \omega_i$, against cavity decay/loss $\kappa$ above the threshold condition. Meanwhile, detuning terms $\Delta\omega$ of each beam around the central cavity resonances are carefully tuned to maintain the phase matching condition. However, this process is complicated by the extra nonlinear phase terms, e.g. SPM and XPM. Previously, more focuses were spotted on XPMs from the pump, since the pump beam is more intense than the other two. For example, it was found out the detuning of pump is always blue during the thermal locking process to overcome the extra nonlinear phase perturbation from the pump's XPM [25]. Here with the growing intensities of succeeding signals and idlers, their own NPMs besides the pump's can no long be neglected.

$$\frac{dA_s}{dt} = (-\kappa + i\Delta\omega_s)A_s + \sqrt{2\kappa_e}A_s^{in} + i\gamma \left( \underbrace{A_s|A_s|^2}_{SPM} + \underbrace{2A_s|A_p|^2}_{XPM} + \underbrace{2A_s|A_i|^2}_{XPM} + \underbrace{A_pA_pA_i^*}_{FWM} \right) \quad (1)$$

$$\frac{dA_i}{dt} = (-\kappa + i\Delta\omega_i)A_i + i\gamma \left( \underbrace{A_i|A_i|^2}_{SPM} + \underbrace{2A_i|A_p|^2}_{XPM} + \underbrace{2A_i|A_s|^2}_{XPM} + \underbrace{A_pA_pA_s^*}_{FWM} \right) \quad (2)$$

Another key argument to support the existence of signals and idlers' NPMs lies on the steady state analysis of OFCs. At the steady state, we can see the balance between input pump's gain and the cavity losses of each comb line at their resonances from Eqns. (1-2). To reach such delicate equilibriums among the single pump and many comb lines, NPMs play an important part in detuning these comb lines from their cavity resonances in order to maintain the balance between the gain and the loss, meanwhile, fulfilling both energy and momentum conservations for FWMs. However, nonlinear phase corrections from pump's NPMs are not enough to regulate all the comb lines simultaneously; NPMs from each comb line must participate into the process. OFC lines grow from initial zero value (gain > loss) to a steady state level (gain=loss), this variable gain must depend on comb lines' own intensities, which helps stabilizing gains through NPMs. A similar process occurs in a laser, for example, a single mode laser at the threshold indicates the balance of the laser gain and the cavity loss. Further increase of gain will not directly lead to a second lasing mode, instead, nonlinear gain saturation or nonlinear detuning of cavity resonances ensures the rebalancing of gain and loss, leaving single mode operation. Hence, NPMs here indeed play a major role in stabilizing lasers or OFCs.

To show the true nature of signals' and idlers' NPMs apart from pumps' is a challenging task, we experimentally demonstrate the transient dynamics during OFCs' formation caused by NPMs near the threshold of parametric oscillations, by temporally modulating the pump beam as shown in Fig.2. Here the pump beam with a power of 5 mW and at 1550 nm wavelength is thermally locked to a high-Q ($\sim 1.0 \times 10^7$) microsphere cavity made of silica through a taper fiber to generate OFC with only five come lines for simplicity. The pump laser is temporally modulated using a square wave at a repetition rate of 100 kHz and 90% duty ratio, which ensures the thermal stability and allows us to monitor the transient dynamics at the same time. Figure 2a shows the key findings of this work: the NPMs help stabilizing OFC formations during the transient regime. The gated gain from parametric FWMs induced by the pump beam gradually increases in each frame of cycles through cavity enhancement effect, when the pump beam is approaching the center of its resonance according to external laser detuning. In this manner, we can obtain a controllable quasi-stable gain while not bothered by other unstable problems, e.g. loss of thermal locking of the pump.

As shown in Fig. 2, the signal comb line starts to grow right after the gated gain, gradually reaching a steady-state level and stabilizing itself for a relatively small gain. For a larger gain, the signal rapidly rises and quickly falls right afterwards, forming a "spike". The same effect can be observed to the idler side simultaneously. These self-evolving processes indeed result from NPMs: the FWM parametric gains exceed the cavity losses, generating signals and idlers. The growing power of signals and idlers effectively detunes their own resonances through NPMs, reducing the gain until the steady state level where the gain equals to the loss again. This gain reduction has to depend on its own intensity, not pump's, due to nonlinear self-detuning caused by their own NPMs rather than the pumps' (constant during the process) as shown in Equ.1&2. Much like in a laser operation when nonlinear gain saturation kicks in above the threshold condition to stabilize the laser. In a similar way, if the starting gain is far more intense, the signal firstly rise to a peak where the effective gain has went far below the cavity loss (Fig. 1d), this leads to a sudden intensity drop to form a spike, exactly following the laser spiking mechanism[19]. This conclusion is justified by our numerical simulations (Fig. 2b), where this self-stabilizing and the spiking effect only present themselves if the NPM terms from signals and idlers are included in the numerical codes.

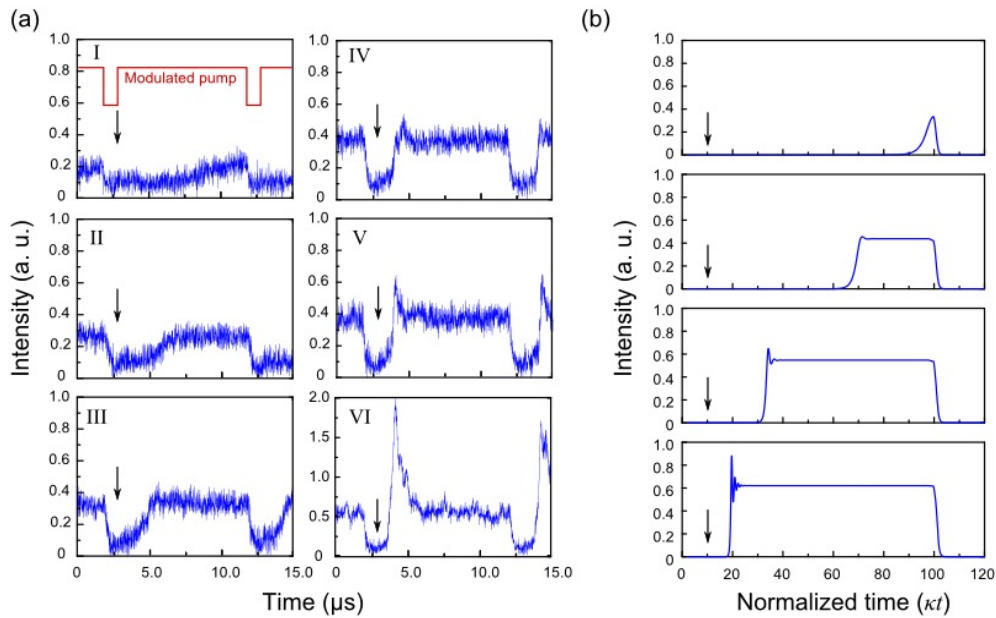

Figure 2. Experimental observations of gain spiking of OFCs. (a) Time trace of signal as pump is gradually tuned towards the resonance center, enhancing the gain. The signal rises faster with the increasing of the gain. A gain spiking of signal is formed in the large gain regime. The arrows show the turn-on points of the pump. (b) Numerical simulations of signal dynamics at the presentence of NPMs with increasing gain from top to bottom.

As shown above, NPMs are crucial in stabilizing OFCs during transition. Furthermore, in order to reveal the true nature of XPMs between signals and idlers since they are almost evenly intense inside microcavities given by the parametric character of FWMs, we must introduce an external perturbation to break this symmetry. Here in our experiment, a second external coherent laser is injected to a stable OFC (constant pump) by sweeping its frequency (constant power, without modulation) around one comb line's spectral windows. Its power is at the similar level of the signal output from the taper fiber, far below the pump's power to avoid disturbing the original OFCs. As the external laser is tuned around the central signal line, these two coherent lasers initiate interference beating between themselves, which is purely *linear effect*. On the other idler spectrum, when the external probe is small, not much intensity fluctuation is observed except at the center of idler line, where an intensity spike exhibits (Fig. 3a), indicating extra parametric gain brought by the external probe through the cavity enhancement process. More interestingly, when the pump's intensity increases, a certain amount of oscillations appears on both wings of central idler line. This is because the external probe now induces its own idler through optical parametric amplification (OPA) much like double-pumped OFCs [28], which *linearly* interferences with the original idler comb line. However, such interference beatings diminish in both signal & idler windows near the original comb lines' central locations, implying that two parametric processes, i.e. the OFC and the OPA of the probe, have emerged into one. The width of this regime is effectively measurement of OFC's gain width, proportional to the pump as shown in Ref. [9]. Such no-beating regime shows that OFCs can occur in a wider spectrum with an external perturbation, where this external probe now can govern the signal line's frequency. During this process, NPMs especially XPMs must participate to regulate nonlinear phase of the signal and idler lines since NPMs from the pump is maintained at a constant level. This regime creates a unique platform in the following section to study the effect of XPMs, since we gain the access of controlling the signal comb lines through this manner. This nonlinear beating phenomena will also be useful for OFCs with multiple pumping scheme [28,29].

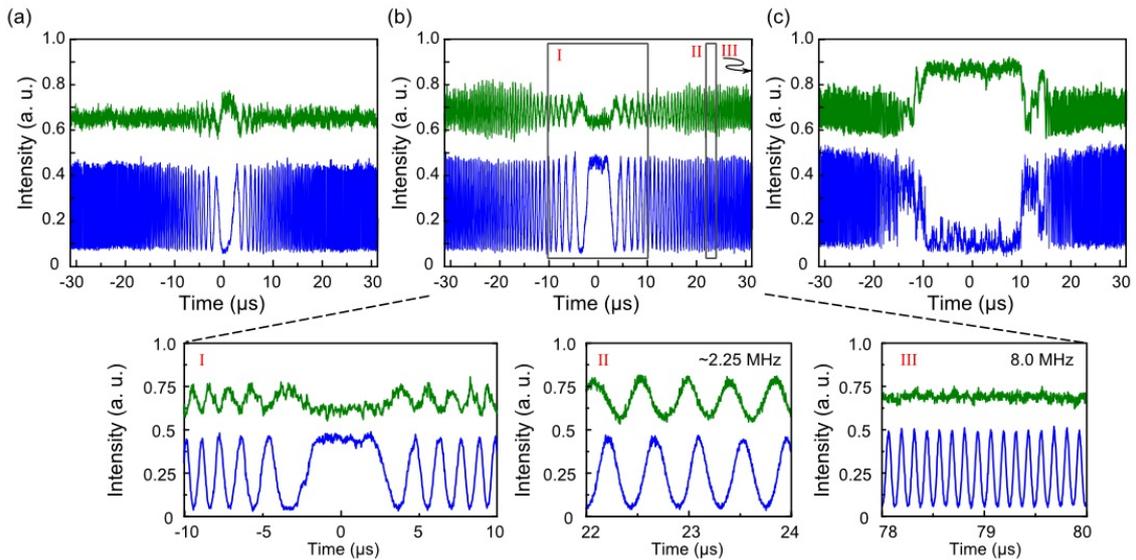

Figure 3. Nonlinear optical beating of OFC and an external probe in the microcavity. The beating of signal (linear beat) was achieved by sweeping another wave across the signal comb line at 1539.3 nm. The pump @1550.6 nm is maintained during the sweeping at constant power without modulation. OFC is the same as Fig. 1b. As the pump power increases from (a) to (c), the coalescence of the signal waves when their frequencies are close enough is clear. A beat note was observed at the idler (1562.0 nm) spectrum, but beating vanished nonlinearly in the center of resonance. The scan speed is 0.1 MHz/us.

To directly demonstrate the effect of XPMs, we settle the external probe in the no-beating regime near the center of resonance. In this manner, we once again adopt the modulation technique by modulating the probe laser to observe OFC's transient behaviors. As shown in Fig. 4, the idler intensity's variation can depend on the modulation depth of the signal. When the inner-cavity signal modulation is small, the idler's waveform is dramatically distorted, behaving like a slow rise-and-fall in Fig. 4a. As a contrast, with increasing of modulation depth of the inner-cavity signal, such rising/falling edges becomes sharper and more instant, almost imprinting the signal's square waveform. Note that, the inner-cavity signal modulation (red curves in Fig. 4, which are difficult to obtain experimentally in transient regime) is reverse to the external probe's since the signal output's combines the residue of the probe and the out-coupled inner-cavity signal. The cause of action is very similar to gain detuning processes with the pump modulation as mentioned above : the XPM from the signal causes the effective detuning of the idler's resonance, breaking the balance of the gain and the loss. Right after the sudden change of rising/falling edges of signals', the idler tends to re-stabilize itself after this sudden detuning caused by XPM. Certainly, larger modulations will lead to wider unbalanced gain and loss, giving a quicker re-stabilization in Fig. 4c. The observation here of asynchronous or hysteresis behaviors between the signal and the idlers may enable an important applications in all-optical multiplexing, but in an omnidirectional way, i.e. exchanging information between the signal and the idler's channels in a one way manner, which has been highly demanded in information science [30]. In a similar way, a non-reciprocal transmission can also be obtained by exploring this nonlinear asymmetry in a single cavity [31]. More generally, this nonlinear broken symmetry may deeply link to the unsolved issue in OFCs, i.e. asymmetrical combs, where previous works relate such asymmetry of comb lines to linear factors such as loss, dispersion [21]. Here, we believe such nonlinear NPMs may also play an important role in generating asymmetry OFCs, which may open up a new avenue in all-optical signal processing such as non-reciprocal multiplexing.

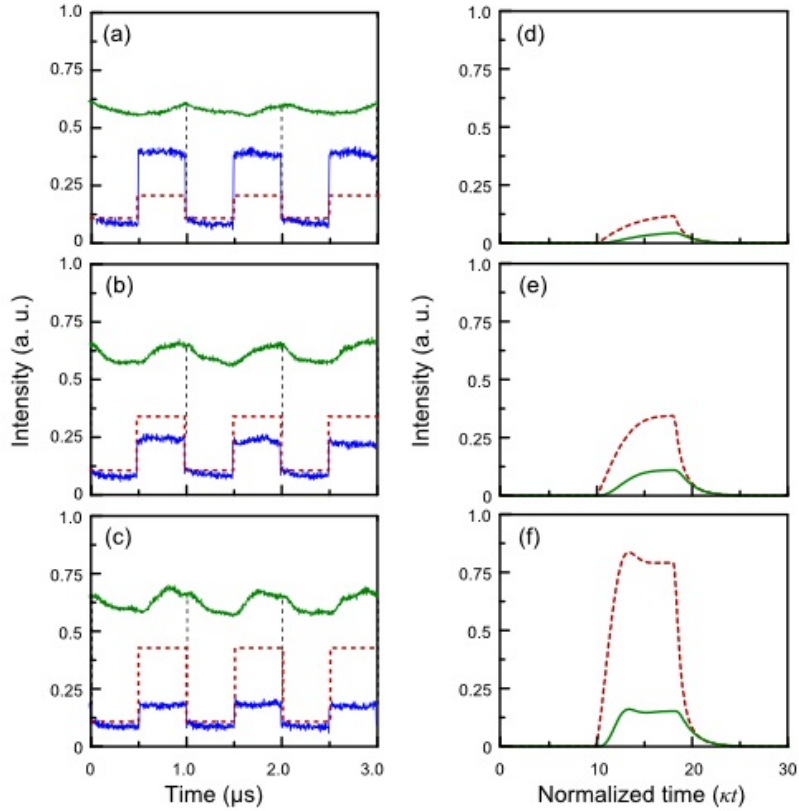

Figure 4. NPMs between signals and idlers of OFCs. (a-c) experimental observation, (d-f) numerical simulations of idler re-stabilization through XPMs of signal modulation. The green, blue, red curves are for idler, signal output, inner-cavity signal, respectively.

At last, we would like comment on the implication of the current work. Here we reveal a crucial but often omitted nonlinear effect of NPMs during OFC generation. The significance of NPMs become more and more profound when OFCs explore new era in ultrashort pulsing operation, wider spectrum generation and other important experimental scheme as multi-pumping, microwave generation through optical beating. Obviously, NPMs may provide the key solution in all areas to stabilize OFCs. For example, one remaining challenge is to achieve self-referenced frequency comb generators on a chip scale microcavity. The difficulty is to generate octave-spanning combs over a wide spectrum, though previous attempts in improving the optical quality factors and flattening dispersion has yet produced such OFCs. We expect NPMs discovered in this work may offer a key answer to this problem.

**Acknowledgements:**
National key research and development program (Grant No. 2016YFA0302500); Natural Science Foundation of China (Grant No. 11674228 No. 11304201, No. 61475100); National 1000-plan Program (Youth); Shanghai Scientific Innovation Program (Grant No. 14JC1402900); Shanghai Scientific Innovation Program for International Collaboration (Grant No. 15220721400).